\documentstyle[pra,aps]{revtex}
\begin{document}
\draft
\font\Bbb =msbm10  scaled \magstephalf
\def\id{{\hbox{\Bbb I}}}

\title{From limits of quantum nonlinear operations
to multicopy entanglement witnesses and state spectrum estimation}

\author{Pawe\l{} Horodecki\cite{poczta2}}

\address{Faculty of Applied Physics and Mathematics\\
Technical University of Gda\'nsk, 80--952 Gda\'nsk, Poland}

\maketitle

\begin{abstract}
The limits of nonlinear in quantum mechanics are
studied. The impossibility of physical implementation of the
transformation $\varrho^{\otimes n} \rightarrow \varrho^{n}$
in quantum mechanics is proved.
For sake of further analysis the simplest notion
of structural completely positive
approximation (SCPA) and structural physical approximations (SPA)
of unphysical map are introduced.
Both always exist for linear hermitian maps and can
be optimised under natural assumptions.
However it is shown that some intuitively natural SPA
of the nonlinear operation $\varrho^{\otimes 2} \rightarrow \varrho^{2}$
that was already proven to be unphysical is impossible.
It is conjectured that there exist no SPA of the operation
$\varrho^{\otimes n} \rightarrow \varrho^{n}$ at all.
It is pointed out that, on the other hand, it is
physically possible to measure the trace of the second power
of the state $Tr(\varrho^{2})$ if only two copies of
the system are available. This gives the
interpretation of one of  Tsallis entropy
as mean value of some ``multicopy'' observable.
The (partial) generalisation of this idea
shows that each of higher order Tsallis entropies
can be measured with help of only two multicopy observables.
Following this observations the notion of
multicopy entanglement witnesses is defined and
first example is provided. Finally, with
help of multicopy observables simple method of spectrum
state estimation is pointed out and discussed.
\end{abstract}


\section{Introduction}
The limits of nonlinear operations within quantum
mechanics is an interesting question.
It has been shown that, \cite{Gissq} for example the operation
\begin{equation}
\varrho \otimes \varrho \rightarrow
\left[ \begin{array}{cc}
	  \varrho_{11}^{2} & \varrho_{12}^{2}	\\
	  \varrho_{21}^{2} & \varrho_{22}^{2}  \\
       \end{array}
      \right ],
      \label{tran}
      \end{equation}
can be performed with the finite probability
by means of quite simple network with two copies of $\varrho$ as an input.
On the other hand the limits for other nonlinear operations has been
shown resulting in ``no-disentanglement'' rule in quantum mechanics
\cite{Danny}, \cite{TD}. In this work we want to show
both further limits and advantages of nonlinear transformations
in context of quantum entanglement theory.

It can be easily seen that if the state in (\ref{tran}) is
diagonal then we get the square of it.
However to get this we have to know the
eigenvectors of the state. It is interesting that if square power
were possible for unknown state we would be able to distill
entanglement from large classes of the states
with little previous information about them
\cite{Distillunknown}.
To some extent it would be similar to
the situation in the compression protocol of Ref. \cite{Jozsa}
where no measurement of the source state is needed
if one of its parameters (entropy) is known.
We shall show that it is impossible
to produce any power of the state
if we do not know its eigenvectors.
Namely it is impossible to perform the operation providing ``n-th''
power of the {\it unknown} state from $n$-copies of it.

However, one can weaken requirements:
sometimes it is impossible to perform some operation
but is is possible to perform it {\it approximately}.
So one can try to perform such approximation.
The well known example are cloning operation \cite{Buzek1,Buzek2,Werner} and
transposition (or universal NOT gate) \cite{Werner1}
and  ``two-qubit fidelity'' map \cite{Winter}.
Recently more careful study of approximations of one qubit maps
has been carried out \cite{Oi1,Oi2}.

It is natural to expect that physical approximations
of unphysical maps could help in solving physical problems in
general. To study this we use the notion of structural completely
positive approximation (SCPA) and structural physical approximation
(SPA) of the unphysical operation.
Those are very restrictive approximations -
the key feature of them
is that they always have the direction
of generalised Bloch vector of the output state
the same as the output state of the original unphysical map.
Only the length of the vector is rescaled by some factor.

We point out  that SCPA and SPA always exist for linear unphysical hermitian
maps. We also prove that there is natural optimisation giving
the best SPA. Hovever, we give the proof that the most natural
trace preserving SPA of unphysical map
$\varrho^{\otimes 2} \rightarrow \varrho^{2}$ is impossible.
We rise the question of whether SPA of such nonlinear maps in general.

However, as we shall see,
the trace of the $n$-th power of the state i. e. the value $Tr(\varrho^{n})$
can be simply measured if $n$  copies of the system are available.
We show how to do it in practice by means of generalised
``swap'' operator. We show several interesting applications of
that fact. Namely one can apply what we propose to
call {\it multicopy observables} of the system: mean value of such observables
is measurable if joint measurement on several copies of the system
in the same state is achievable. We point out that, following the latter,
that the Tsallis entropy $S_{2}$ can be
treated as twocopy observable.
Further all Tsallis entropies $S_{q}$ of natural index
$q=n>2$ can be measured with help of only two
multicopy observable. Then applying the separability
conditions in terms of entropic inequalities
(initiated in \cite{vonN}, developed in
\cite{alphaPLA,pra,Cerf,Cerf1,Smolin,Barbara} and completed in an elegant
way in \cite{Wolf}) we show that some Tsallis
entropic separability conditions (equvalent to quantum Renyi ones)
can be checked directly (or almost directly)
with help of multicopy observables.
This is remarkable as only finite (small) number
of copies in joint measurement is required.
Finally we point out how to estimate the spectrum of unknown
state using the idea of multicopy observables.
This observation is in a sense complementary to the
optimal procedure of Ref. \cite{Spectrum}.
The advantage is that collective measurements
on only finite (not more than the dimension of single
system Hibert space) number of copies is needed.

The paper is organised as follows:
In section II we prove the following ``no go''
result: if the state $\varrho$ is
unknown than the operation $\varrho^{\otimes n} \rightarrow
\varrho^{n}$ is impossible.
In sect. III we provide  general
idea of structural completely positive approximation (SCPA)
and its slight modification
 -  structural physical approximation (SPA).
In sect. IV we
show that those approximations always exists
for any hermitian map. We also show that the most
natural SPA is optimal.

In section V we apply the concept
of SPA to nonlinear quantum maps showing that
the intuitively most natural tracepreserving
SPA of unphysical operation
$\varrho^{\otimes 2} \rightarrow \varrho^{2}$
does not exist.
The general ``no go'' conjecture is also formulated.

In section IV we investigate further possiblity
of direct measurement of nonlinear parameters.
We introduce the notion of ``multicopy
observable'' and show how Tsallis entropies
can be measured with help of such observables.
We utilise entropic separability
criteria from the literature of the subject
we introducing the notion of {\it multicopy entanglement
witness} i. e. such observable detecting
entanglement of $\varrho$ that is measured jointly
on the several copies of the system.

Finally in short Section V we poit out  a simple method of estimation
of spectrum of unknown state defined on ${\cal C}^{m}$
which requires collective measurement of some number
of copies, but only estimation of $2m-3$ parameters
is needed (instead of $m^{2}-1$ due to usual quantum tomography)
that are mean values of some multicopy observables.

\section{Proof of impossibility of operation
$\varrho^{\otimes n} \rightarrow \varrho^{n}$}

Consider an arbitrary quantum state defined on ${\cal C}^d$
space. We shall show that there is no quantum transformation
of the kind
\begin{equation}
\Lambda(\varrho^{\otimes n}) \rightarrow \varrho^{n}
\label{potega}
\end{equation}
which works for {\it unknown} quantum state.
Let us note first that such operation would be
{\it probabilistic} i. e. it would give the required output with
the probability $p=Tr(\varrho^{n})$ depending on the input state. Suppose that such
operation existed. Then
because of complete positivity
it would be of the form
completely positive map $\Lambda (\varrho)= \sum_{i=1}^{N}V_{i} \varrho
V_{i}^{\dagger}$.
So we would rewrite (\ref{potega}) as follows
\begin{equation}
\sum_{i=1}^{N} V_{i} \varrho^{\otimes n} V_{i}^{\dagger}=
\varrho^{n}
\end{equation}
Taking trace of both sides of the above
we get that there would exist the {\it positive}
state independent operator
$A=\sum_{i=1}^{N} V^{\dagger}_{i}V_{i}$,
with the property
\footnote{If $A, B$ are hermitian operators
then the notation $A \leq B$ means that
for any vector $|\psi\rangle$ one has
$\langle \psi| A|\psi\rangle \leq \langle \psi| B|\psi\rangle$.}
$0\leq A \leq I$ such that
\begin{equation}
Tr(A\varrho^{\otimes n})=Tr(\varrho^{n})
\label{2}
\end{equation}
for {\it any} state $\varrho$.
In particular for any pure projector $P_{\phi}=|\phi\rangle \langle \phi|$
corresponding to normalised vector $|\phi\rangle$
we would have $Tr(AP_{\phi}^{\otimes n})=Tr(P_{\phi})=1$.
But, because all eigenvalues of $A$ belong
to the interval $[0,1]$, this means that any vector of
the form $|\Psi \rangle=|\phi \rangle^{\otimes n}$ must
be an eigenvector of $A$. However
all $|\Psi \rangle$-s of that form span the completely symmetric
subspace ${\cal H}_{SYM}$ of $({\cal C}^{d})^{\otimes n}$.
Thus the subspace is an eigenspace of $A$ corresponding to
eigenvalue $1$. Now for any $\varrho$ the support of
the operator $\varrho^{\otimes n}$ belongs to the
${\cal H}_{SYM}$. So $A \varrho^{\otimes n}= \varrho^{\otimes n}$
which means that LHS of the equation (\ref{2})
should be {\it always} $1$ which clearly leads
to the contradiction for any $\varrho$ which is not pure.

\section{Structural physical approximations
of unphysical maps: the concept}
\subsection{Definition}
\label{definitionSCPA}
Consider now the possibility of obtaining
the completely positive approximation
$\overline{\Theta}$ of the physically impossible operation $\Theta:
{\cal B}({\cal C}^{d}) \rightarrow {\cal B}({\cal C}^{d'})$.
We shall require that the following ``error'' operator
\begin{equation}
\Delta(\varrho)=\overline{\Theta}(\varrho)-\gamma(\varrho)\Theta(\varrho)
\label{delta}
\end{equation}
satisfies the invariance condition
\begin{equation}
\Delta(\varrho)=\delta(\varrho)I
\label{invariance}
\end{equation}
for identity operator $I$ on ${\cal C}^{d'}$
and some scaling parameters $\gamma(\varrho)\geq 0$,
$\delta(\varrho)\geq 0$ depending in general on $\varrho$.

More precisely we require what
can be represented by the following proposed definition

{\it Definition 1.- The structural completely
positive approximation (SCPA) of unphysical map $\Theta$
is any completely positive operation of the form
\begin{equation}
\overline{\Theta}(\varrho)=\delta(\varrho)I+\gamma(\varrho)\Theta, \ \
\label{forma}
\end{equation}
with the functions $\delta, \gamma \geq 0$
and $\gamma$ strictly positive for all $\varrho$
such that $\Theta(\varrho)>0$.
The structural physical approximation
(SPA) of unphysical map $\Theta$ is such SCPA $\overline{\Theta}$
that for any state $\varrho$ $Tr(\overline{\Theta}(\varrho))\leq 1$
i. e. that can be implemented experimentally.}

{\it Remark .- } The the first SCPA (and as we shall
see also SPA) was optimal NOT gate from Ref. \cite{Werner1}.
The approximated cloning machine
(\cite{Buzek1,Buzek2,Werner}) was not because
one the output of the machine
is entangled while according to the present
definition it should be separable.

The essence of SCPA of any $\Theta$ is that it (i) is completely
positive and (ii) {\it  keeps the structure of the output}
of the unphysical operator $\Theta$.
In other words for any argument
the direction of generalised Bloch vector of the output
matrix
is the same as as the direction of the output matrix
of the original unphysical map.
The output is however ``shrinked'' by
factor $\gamma$ (a kind of ``Black Cow'' factor,
see \cite{Werner1}) and the additional portion (quantified
by $\delta$) of completely random noise is admixed.
The SPA is such SCPA that can be probabilistically
implemented in lab (see Appendix).
Note that for finitedimensional systems some
SPA can be obtained form nonzero SCPA by normalisation
\begin{equation}
\overline{\Theta}_{SPA}=t^{-1}\overline{\Theta} \ \ ,
t\equiv\mathop{\mbox{max}}\limits_{\varrho} Tr[\overline{\Theta}(\varrho)].
\end{equation}
where strict positivity of $t$
is given by complete positivity of nonzero SCPA $\overline{\Theta}$.

Now one can ask: what is the optimal approximation in sense of the
above definitions ? Let us recall that the question of optimality
was frequently posed in context of approximate cloning machines
etc. In this context we can discuss the proposal of the
following notion of optimality:

{\it Definition 2 .- The best (or the optimal) SPA
of $\Theta$ is such SPA map $\overline{\Theta}_{opt}$
that (i) minimizes the ratio
$\delta(\varrho) / \gamma(\varrho)$ for
any $\varrho$ in (\ref{forma})
(ii) maximizes $Tr[\overline{\Theta}(\varrho)]\leq 1$
over all SPA-s satisfying property (i).}

This would mean that any other SPA
has for any $\varrho$ the ratio greater than
the one of the best SPA.
The point (ii) would have similar meaning.
The idea of the above notion would be that the ratio of
the noise to the approximated map $\Theta$ in (\ref{forma}) would
``prefer'' the latter as much as possible.
For the best SPA one requires in addition
as much probability of implementation
as possible (see Appendix).

The problem with the above notion of optimality
is that it is difficuilt to check whether such optimal maps
always exist.
In particular it is quite probable that the best SPA
may not exist in some peculiar cases
(for example when $\Theta$, $\delta$ or $\gamma$ are not continuos).
However, as we shall see below, for linear hermitian maps they always exist
and under natural conditions they can be
easily optimiesed.

\section{Physical approximations of hermitian maps}
\label{BSCPA}
\subsection{Natural construction}

By hermitian we shall regard any map
that preserves hermicity property. In what follows
we shall briefly prove very simple

{\it Proposition 1 - For any hermitian linear map
$\Theta: {\cal B}({\cal C}^{d}) \rightarrow
{\cal B}({\cal C}^{d'})$ there exists SPA
$\overline{\Theta}$ defined by
\begin{equation}
\overline{\Theta}_{a}=t^{-1}(aI_{d'}+\Theta)
\label{formula}
\end{equation}
with
the parameter
$a \geq \lambda d\equiv max[0,-\lambda'd]$
where $\lambda'$ is the minimal eigenvalue of
the operator $[\id \otimes \Theta](P_{+})$.
Here $P_+=|\Psi_+\rangle\langle \Psi_+|$
corresponds to the ``isotropic'' maximally entangled
$d \otimes d$ state $|\Psi_+\rangle=\frac{1}{d}
\sum_{i=1}^{d} |i\rangle|i \rangle$,
$t\equiv\mathop{\mbox{max}}
\limits_{\varrho} Tr([aI_{d'}+\Theta](\varrho))$}.

In the above one must remember that $I_{d'}$
stands for identity matrix acting on ${\cal C}^{d'}$
and we ideintify with this matrix ``maximal noise''
map that replaces every operator with $I_{d'}$.
Note also that $[\id \otimes I_{d'}](P_{+})=\frac{1}{\sqrt{d}}I_{d} \otimes I_{d'}$.
Subsequently we shall omitt indices at $I_{d'},I_{d}$.

The complete positive character of approximation (\ref{formula})
follows immediately from the well-known
fact that linear hermitian operation $\Lambda$ is completely possitive
iff $[\id \otimes \Lambda](P_{+})$ has nonegative
spectrum. The approximation is also the most natural,
as it does not involve any nonlinear function of $\varrho$
and relies on adding only ``backgroud noise''
to the original map.

\subsection{Question of optimality}
Let us call any hermitian
map {\it nontrivial}  if it is not of the form
$\Lambda(\varrho)=c(\varrho)I$ (with $c(\varrho)$ being some real function)
i. e. it does not map all states
into (possibly rescaled) identity matrix.
Let also call SCPA and SPA {\it regular}
if it has the shrinking function $\gamma$
continuous. We have the following

{\it Proposition 2 .- From all regular SPA-s
of nontrivial linear hermitian map $\Theta$
the approximation (\ref{formula}) with
minimal $a$ ($a=\lambda/d$) is the best one.}

{\it Proof.-}
Let us recall that the spaces ${\cal B}({\cal C}^{d})$,
${\cal B}({\cal C}^{d'})$
are Hilbert spaces (known as Hilbert-Schmidt spaces)
with a scalar product
$\langle A |B\rangle =Tr(A^{\dagger}B)$.
Let us take {\it support} of original $\Theta$.
This is that subspace of ${\cal B}({\cal C}^{d})$
on which $\Theta$ does not vanish.
In this subspace we can find the orthonormal basis
consisting on $r$ elements $\{ X_{i} \}_{i=1}^{r}$.
There exists its complement called {\it kernel} of
$\Theta$ represented by the set of $k$ elements $\{ X_{i} \}_{i=r+1}^{k} \}$
so that the domain ${\cal B}({\cal C}^{d})$ is spanned
by all $r+k=d^{2}$ elements.
Because linearily independent (even orthonormal)
elements of support $X_{i}$ must be mapped into linearily independent ones
we get that the original map is of the form
\begin{equation}
\Theta(A)= \sum_{i=1}^{r}
 Y_{i} Tr[X^{\dagger}_{i} A]
\label{exp}
\end{equation}
on arbitrary $A$
with linearily independent operators $Y_{i}\in {\cal B}({\cal C}^{d'})$.
Because $\Theta$ is nontrivial at least
one of $Y_{i}$, say $Y_{i_{0}}$, must be linearily independent
on operator $I$. An arbitrary operator $X$
in the domain of $\Theta$ is of the form
$X=\alpha X_{i_{0}} + \beta X'$
where $X'$ is linear combination of
$X_{i}$-s orthogonal to $X_{i_{0}}$.
Let us consider the action of $\overline{\Theta}$ (\ref{forma})
(which is linear) on $X$.
Form linearity of $\Theta$, $\overline{\Theta}$
and analysis of the coefficient at $Y_{i_{0}}$ in
the corresponding expansion we get $\gamma(\alpha X +
\beta X_{i_{0}})=\gamma(X_{i_{0}})$ for any nonzero $\alpha$.

From regularity of $\overline{\Theta}$ (which means
continuity of $\gamma$) we get that $\gamma$ is
a constant fuction equal to $\gamma(X_{i_{0}})$.
Note that it was not obvious because
$\gamma$ (and also $\delta$) in principle
could be nonlinear functions of the state.
Because of hermicity of $\overline{\Theta}$ constant $\gamma$
is represented by real number that we shall denote
also by $\gamma$. This immediately implies linearity
of function $\delta$. Hence, because of
Riesz theorem applied to Hilbert-Schmidt space
$\delta$ is uniquely determined by some hermitian
operator $D$ in the following way: $\delta(\varrho)=Tr(D\varrho)$.
Now the complete positivity of $\overline{\Theta}$
results in condition $\delta(\varrho) \geq \gamma \lambda d$
where $d$ is a dimension of Hilbert space
and $a \geq \lambda$ is defined as in Proposition 2.
Minimisation of the rate $\delta/\gamma$ results here
immediately in (i) minimal value of $a=\lambda$,
(ii) constant character of $\delta$ (it is simply a number)
(iii) the equality

\begin{equation}
\delta=\gamma \lambda d
\end{equation}

This results in  $\gamma$-parameter
family SPA maps of the form $\overline{\Theta} \equiv \overline{\Theta}_{\gamma}=
\gamma (\lambda d I + \Theta)$.
On the other hand to get ``the best'' SPA we have
to maximise the value $Tr(\overline{\Theta}_{\gamma}(\varrho))$
under the condition $Tr(\overline{\Theta}_{\gamma}(\varrho)) \leq 1$.
If for original $\Theta$
one defines $\alpha_{\Theta}$ as the value of
$Tr[\Theta(\varrho)]$ maximised over all states $\varrho$
($\alpha_{\Theta}=\mathop{max}\limits_{\varrho}Tr[\Theta(\varrho)]$) then
we get immediately
(i) optimal $\gamma=\frac{1}{\lambda dd'+\alpha_{\Theta}}$
(ii) optimal
$\delta=\frac{\lambda d}{\lambda dd'+\alpha_{\Theta}}$ and
the best SPA:
\begin{equation}
\overline{\Theta}_{opt}=\frac{\lambda d}{\lambda dd'+\alpha_{\Theta}} I +
\frac{1}{\lambda dd'+\alpha_{\Theta}} \Theta
\label{bSPA}
\end{equation}
One concludes the proof
by checking
that the above is identical with
(\ref{formula}) after putting $a=\lambda d$.
\subsection{Interpretation and example}
There is a simple interpretation of the optimal formula
(\ref{bSPA}) if $\alpha_{\Theta} \neq 0$.
To get the best SPA $\overline{\Theta}$ one need
the following two operations

(i) rescale given linear hermitian $\Theta$
by taking $\Theta_{1}=\alpha^{-1}_{\Theta} \Theta$,
($\alpha_{\Theta}=\mathop{max}\limits_{\varrho}Tr[\Theta(\varrho)]$).
(Then $\Theta_{1}$ already satisfies $Tr[\Theta_{1}(\varrho)] \leq 1$.)

(ii) take the following convex combination
\begin{equation}
\overline{\Theta}_{opt}=p_{*}\frac{I}{d'}+(1-p_{*})\Theta_{1}
\end{equation}
with probability $p_{*}=
\frac{\lambda dd'\alpha_{\Theta}^{-1}}{\lambda dd'\alpha_{\Theta}^{-1}+1}$
and $\lambda=max[0,-\lambda']$ where $\lambda'$ is the
minimal eigenvalue of the operator $[\id \otimes \Theta](P_+)$.

In the case of tracepreserving maps
(or in general all maps that satisfy $\alpha_{\Theta}=1$)
the above protocol gives $\Theta_{1}=\Theta$
so the only step (ii) taking convex combination is important.
This means that in those cases to
get optimal SPA one have to perform
probabilistic mixture of $\Theta$
with depolarising channel $\Lambda_{dep}:
{\cal B}({\cal C}^{d})\rightarrow {\cal B}({\cal C}^{d'})$
defined by $\Lambda_{dep}(\varrho)=\frac{I}{d'}$
for any $\varrho$.
In particular we have the

{\it Example .-} Consider the best SPA  of the transposition
map $T: {\cal B}({\cal C}^{d}) \rightarrow
{\cal B}({\cal C}^{d})$ which transposes
the matrix $[T(A)]_{mn}=A_{nm}$.
Applying the prescription above for $\Theta=T$
we get
$[\id \otimes T](P_+)=
\frac{1}{d} V$ where $V$ is a ``flip'' or ``swap''
operator \cite{Werner0}. As $V$ has spectrum $\pm 1$
this gives $\lambda=-\frac{1}{d}$,
$\alpha=1$ which results in
the optimal parameter $p^*=\frac{1}{d}$.
In this way we get tracepreserving SPA
which is the following quantum channel
\begin{equation}
\overline{T}: \varrho \rightarrow {d \over d+1}\frac{I}{d} + \frac{1}{d}\varrho^{T}
\end{equation}
or $\overline{T}=\frac{d}{d+1}\frac{I}{d}+\frac{1}{d}T$.
It has already been realised
as a byproduct of optimal quantum cloning machines
\cite{Buzek1,Buzek2,Werner} and is closely
related to universal quantum NOT gate \cite{Werner1}.

\section{Some limit of approximations of nonlinear maps}
What about our map making the power of
the state? Can we approximate it in a way
described above? We conjecture
that it is impossible in general.
We shall consider the case which seems to be natural to look at.
Namely consider the hypothetical map
$\Lambda_{\mbox{?}}: {\cal B}({\cal C}^{d} \otimes {\cal C}^{d})
\rightarrow {\cal B}({\cal C}^{d})$ defined
in the form
\begin{equation}
\Lambda_{\mbox{?}}(\varrho \otimes \varrho)
=(1-Tr(\varrho)^{2}) I/d + \varrho^{2}
\label{map2}
\end{equation}
Here we have the tracepreserving property
manifestly taken into account and the error parameter
at the noise depends on the purity of the state
being equal $1-Tr(\varrho^{2})$. Is it possible to
bulit such map? The answer is negative. Namely we have

{\it Proposition 3 .-  For qubit case the map (\ref{map2})
is not completely positive.}

To prove the above statement let us first
observe that putting two copies of pure qubit states
into (\ref{map2}) we get
\begin{equation}
\Lambda_{\mbox{?}}(|\phi\rangle \langle \phi| \otimes
|\phi\rangle \langle \phi|)= |\phi\rangle \langle \phi|.
\label{qubit1}
\end{equation}
Taking this into account and
putting two copies of mixed states we get
\begin{equation}
\Lambda_{\mbox{?}}(|\phi\rangle \langle \phi| \otimes
|\phi^{\perp} \rangle \langle \phi^{\perp}|+
|\phi^{\perp} \rangle \langle \phi^{\perp}|
\otimes |\phi\rangle \langle \phi|)=
\frac{I}{2}
\label{qubit2'}
\end{equation}
for all pairs of orthogonal vectors $|\phi\rangle$,
$|\phi^{\perp}\rangle$.
But the required action of
the map will not change if we compose
it with the symmetrisation as a first step.
After such modification we have
\begin{equation}
\Lambda_{\mbox{?}}(|\phi\rangle \langle \phi| \otimes
|\phi^{\perp} \rangle \langle \phi^{\perp}|)=
\frac{I}{2}
\label{qubit2}
\end{equation}
Suppose that the map is completely positive
i. e. that it is equal to tracepreserving
map of the form $\Lambda_{CP}\sum_{i=1}^{k} V_{i}
\sigma V_{i}^{\dagger}$. Here
$\sigma$ is arbitrary two-qubit state
and $V_{i}: {\cal C}^{2}
\otimes {\cal C}^{2} \rightarrow {\cal C}^{2}$.
The map is tracepreserving so it represents
a quantum channel. Following the equivalence between
quantum states and quantum channels
we can assume that the number $k$ of the
operators $V_{i}$ is equal to $dim ({\cal C}^{2} \otimes
{\cal C}^{2}) \cdot dim({\cal C}^{2})=
4 \cdot 2=8$.
Let us define the following operation
$\tilde{U}: {\cal C}^{2} \otimes {\cal C}^{2}
\rightarrow {\cal C}^{4} \otimes {\cal C}^{2}$
defined as
\begin {equation}
\tilde{U}=\sum_{k=1}^{8}|k \rangle V_{k}
\label{isometry}
\end{equation}
From tracepreserving property of
the map $\Lambda_{CP}$ we immediately see that $\tilde{U}^{\dagger}
 \tilde{U}=I$ where identity
acts on ${\cal C}^{2} \otimes {\cal C}^{2}$
so it is isometry and preserves scalar product of its arguments.
The isometry has the property that
for any $2 \otimes 2$ $\varrho$ partial trace of
the $4 \otimes 2$ operator $\tilde{U} \varrho \tilde{U}^{\dagger}$
with respect to the left subsystem reproduces the
action of the map $\Lambda_{CP}$. Indeed we have
\begin{equation}
Tr_{A}(\tilde{U} \varrho \tilde{U}^{\dagger})=\Lambda_{CP}(\varrho).
\end{equation}
From the conditions (\ref{qubit1}), (\ref{qubit2}) we
see that $\tilde{U}$ satisfies the two following conditions:
\begin{equation}
\tilde{U}|\phi \rangle |\phi\rangle =|\Phi\rangle |\phi\rangle
\label{war1}
\end{equation}
\begin{equation}
\tilde{U}|\phi^{\perp} \rangle |\phi\rangle =|\Psi_{max}\rangle
\label{war2}
\end{equation}
where $\Psi_{max}=\Psi_{max}(\phi)$ is some maximally entangled
vector in space ${\cal C}^{8} \otimes {\cal C}^{2}$.
The state $|\Phi\rangle=|\Phi (\phi) \rangle$
and lives in ${\cal C}^{8}$.
Properties of $\tilde{U}$ implies that it maps
 the standard basis $|0\rangle|0\rangle$,
$|0\rangle|1\rangle$, $|1\rangle|0\rangle$, $|1\rangle|1\rangle$.
into the orthonormal basis $|\eta\rangle |0\rangle$, $|\Psi_{max}\rangle$, $
|\Psi_{max}'\rangle$,
$|\eta'\rangle|1\rangle$ where
$|\Psi_{max}\rangle$, $|\Psi_{max}'\rangle$ are maximally entangled.
Consider now a normalised vector
$|\phi\rangle =\alpha |0\rangle + \beta |1 \rangle$.
Then according to the above we have
\begin{eqnarray}
&&\tilde{U}|\phi \rangle| \phi \rangle=
 \alpha^{2} |\eta \rangle|0\rangle
+\alpha\beta( |\Psi_{max}\rangle + |\Psi_{max}'\rangle) + \beta^{2} |\eta'\rangle
| 1 \rangle= \nonumber \\
&& |\Phi \rangle(\alpha|0\rangle
+\beta| 1 \rangle)
\label{vec1}
\end{eqnarray}
We can act on both sides the linear operator
\begin{equation}
A_{0} \equiv I \otimes \langle 0|
\end{equation}
Note that $A_{0}: {\cal C}^{2}
\otimes {\cal C}^{2} \rightarrow {\cal C}^{2}$.

Application of $A_{0}$ results in equivalence
$ \alpha^{2} |\eta \rangle
+\alpha\beta A_{0}(|\Psi_{max}\rangle + |\Psi_{max}'\rangle)=
 \alpha|\Phi \rangle$.
This leads to
$\alpha |\eta\rangle + \beta |u_{0}\rangle=|\Phi\rangle$
where $|u_{0}\rangle \equiv A_{0}| \Psi_{max} + \Psi_{max}'\rangle$
and $\alpha$ is nonzero. The vector $|\Phi \rangle$ is
normalised so taking $\alpha$ real, $\beta$ real or
purely imaginary we obtain that independently
on value of $\alpha$
\begin{equation}
\alpha^{2}+(1-\alpha^{2})||u_{0}||^{2}
+ x \alpha\sqrt{1-\alpha^{2}}=1
\label{war}
\end{equation}
where $x$ is equal either real or imaginary part of
$\langle \eta |u_{0} \rangle$. This condition (\ref{war})
gives immediately
\begin{eqnarray}
&&||u_{0}||=1 \nonumber \\
&& \langle \eta |u_{0} \rangle=0
\label{ort}
\end{eqnarray}
Consider now the condition (\ref{war2})
again for the vector $|\phi\rangle =\alpha |0\rangle + \beta |1 \rangle $.
Because $\tilde{U}|\phi \rangle| \phi^{\perp} \rangle$
is supposed to be maximally entangled then
the state $A_{0}\tilde{U}|\phi \rangle| \phi^{\perp} \rangle$
living on ${\cal C}^{4}$ has to have the norm ${1\over 2}$ i. e.
\begin{equation}
||A_{0}\tilde{U}|\phi \rangle| \phi^{\perp} \rangle ||^{2}=\frac{1}{2}
\label{norm1}
\end{equation}

Now we act the operation $\tilde{U}$ on the vector $|\phi\rangle|\phi^{\perp}\rangle$
with normalised $|\phi\rangle=\alpha |0\rangle + \sqrt{1-\alpha^{2}}
|1\rangle$.
Rewriting the condition (\ref{norm1}) with help of
conditions (\ref{ort}), and the fact that $\Psi_{max}$ is maximally entangled
lead simply to
\begin{eqnarray}
&&\alpha^{2}(1 - \alpha^{2})
+2\alpha(1-\alpha^{2})^{3/2}{\rm Re}(\langle \Psi_{max}|\eta\rangle |0\rangle)\nonumber \\
&&+2 \alpha^{2}(1-\alpha^{2}) {\rm Re}(\langle \Psi_{max}'|u_{0}\rangle|0\rangle)
-\alpha^{4} + 2(1-\alpha^{2})^{2}=\frac{1}{2}
\end{eqnarray}

This gives the expected contradiciton as $\alpha$ can have
arbitrary value for the interval $[0,1]$.
This concludes the proof that the map (\ref{map2})
and can not be performed within quantum mechanics.

It seems that the following general conjecture is true.

{\it Conjecture .- There is no SPA of the nonlinear
map (\ref{potega}).}

Note that nonexistence of SPA is equvalent to nonexistence
of SCPA as one can be obtained from another by
suitable rescaling procedure.


To conclude the results the section:
according to quantum mechanics
not only the map $\varrho \otimes \varrho \rightarrow \varrho^{2}$
but also $\varrho \otimes \varrho \rightarrow (1 - Tr(\varrho^{2})I/d+\varrho$
is physically impossible.
However it is very interesting that,
as we shall see subsequently, the value of
$Tr(\varrho^{n})$ can be measured in {\it a very simple} way.

\section{Tsallis entropy as ``multicopy'' observable
and multicopy entanglement witnesses}
\subsection{Multicopy observables}
We propose to extend the notion of quantum observable
$A$ to {n-copy observable $A^{(n)}$}.
Suppose that the system state is defined on Hilbert
space ${\cal H}$. Then measurement of $A$ performed on $\varrho$
leads to the mean value $\langle A \rangle \equiv Tr(A\varrho)$.

Here we propose the simple definition

{\it Definition 3 .- Let  $A^{(n)}$ be the hermitian
operator on ${\cal H}^{\otimes n}$. We
interprete it as {\it n-copy observable}
with respect to the single system defined on a single Hilbert space
${\cal H}$  by defining ``mean value''
of $A^{(n)}$ on $\varrho$ as:
\begin{equation}
\langle \langle A^{(n)} \rangle \rangle=\langle A^{(n)}
\rangle_{\varrho \otimes \varrho} \equiv Tr(A^{(n)} \varrho^{\otimes n})
\end{equation}.}

{\it Example 1: ``Swap'' observable.}
Consider the ``swap'' or ``flip'' operator
\cite{Werner} on two-system space
which has the property $V|\Phi \rangle \otimes |\Psi \rangle =
|\Psi \rangle \otimes \Phi \rangle $ for any $\Phi$,
$\Psi \in {\cal C}^{d}$.
It has been shown \cite{Werner} that
\begin{equation}
Tr(V A \otimes B)=Tr(AB)
\label{AB}
\end{equation}
It can be particularily  easy ilustrated on
product pure state
\begin{equation}
Tr(V|\phi \rangle \langle \phi |\otimes |\psi \rangle
\langle \psi |)=
Tr(|\psi \rangle \langle \phi |\otimes |\phi \rangle
\langle \psi |)=Tr(|\psi \rangle \langle \phi |)
Tr(|\phi \rangle \langle \psi |)=| \langle \psi
| \phi \rangle|^2=Tr(|phi \rangle \langle \phi|\psi\rangle\langle\psi|)
\end{equation}
In a similar way if we consider means value of $V$ in state
$\varrho \otimes \varrho$
with $\varrho=\sum_{i}\lambda_{i}P_{\phi_{i}}$.
Here $P_{\phi_{i}}$ is a projector corresponding
to the eigenvector $|\phi_{i}\rangle \langle \phi_{i}|$ then
we obtain
\begin{eqnarray}
&&Tr(V \varrho \otimes \varrho)
=\sum_{i,j}\lambda_{i}\lambda_{j}Tr(V P_{\phi_{i}} \otimes P_{\phi_{j}})=
\nonumber \\
&& \sum_{i,j}\lambda_{i}\lambda_{j}| \langle \phi_{i}|\phi_{j} \rangle|^{2}=
\sum_{i}\lambda_{i}^{2}\equiv Tr(\varrho^{2}).
\label{ro2}
\end{eqnarray}
This reproduces the value implied by general formula (\ref{AB}).
According to the above ``flip'' or ``swap'' $V$ operator can be viewed
as $2-copy$ observable:
the formula (\ref{ro2}) leads to the conclusion
that the value $Tr(\varrho^{2})$ is measurable
iff two copies of the state are available.
From that we get immediately that
the Tsallis entropy $S_{q}^{T}(\varrho)=\frac{1-Tr(\varrho^{q})}{q-1}$
is measurable for $q=2$. Indeed we take the observable $W\equiv I-V$
and $S_{q}^{T}(\varrho)=\langle W \rangle_{\varrho \otimes
\varrho}=Tr(W \varrho \otimes \varrho$).

{\it Example 2: ``Shift'' operation and related mulicopy observables.}
Consider the well-known natural generalisation of the ``swap''
. This is a ``shift'' operation $V_{n}$
which can be interpreted as some cyclic permutation.
It is defined as $V^{(n)} u_{1} \otimes u_2 \otimes ... \otimes u_{n}=
u_2 \otimes ... \otimes u_{n} \otimes u_{1}$.
It is known that (see \cite{Isham})
$Tr(V^{(n)} A_{1} \otimes ... \otimes A_{n})=
Tr(A_{1}... A_{n})$.
Unfortunately is in general not hermitian (which can be
checked directly looking at its decomposition into
``swaps'' $V=V^{(2)}$).
But $Tr(X \sigma)$ for {\it any} operator $X$
and state $\sigma$ can be experimentally
checked by measuring hermitian
(defined as $X_{h}\equiv\frac{1}{2}(X+X^{\dagger})$)
and antyhermitian
(defined as $X_{a} \equiv \frac{i}{2} (X-X^{\dagger}$)) part of
$X$ because
\begin{equation}
 \langle X \rangle= Tr(X \sigma)=
 Tr(X_h \sigma) -iTr(X_a \sigma)=
 \langle X_h \rangle-i \langle X_a \rangle.
 \label{ah}
\end{equation}

Thus one can determine ``mean value'' $Tr(V^{(n)} \sigma)$
of nonhermitian operator $V^{(n)}$
by experimental measurement of only {\it two} hermitian observables.

\subsection{Multicopy entanglement witnesses}
As a separability conditions entropy inequality
was initiated in \cite{vonN} and continued \cite{alfaPLA,pra,Cerf} as
 Renyi entropy analysis in context of separability.
At present we know that any separable state satisfy
the entropy inequality
\begin{equation}
S_{\alpha}(\varrho_{AB})- S_{\alpha}(\varrho_{X}) \geq 0,
X=A,B
\label{alfa}
\end{equation}
they are equivalent to the Tsallis entropy inequalities:
\begin{equation}
S_{q}^{T}(\varrho_{AB}) - S_{q}^{T}(\varrho_{X}) \geq 0,
X=A,B
\label{q}
\end{equation}
The fact that inequalities above are satisfied
byseparable states was proved in steps
(for different parameters $\alpha$)
\cite{vonN,alfaPLA,pra,Cerf,Smolin,Barbara,Lloyd}
with proof completing the result for all $\alpha$-s given in Ref. \cite{Wolf}.
As such they form necessary conditions for separable states.
Now for $q=2$  we easily show that the condition can be measured
directly by the following observable - an example of
what we call {\it multicopy  entanglement witness}.
Consider the two copy state $\varrho \otimes \varrho
\equiv \varrho_{AB}\otimes \varrho_{A'B'}$
on the systems $AA'BB'$
and the following observables
\begin{equation}
W^{A}=V_{(AA'),(BB')}-I_{AA'}\otimes V_{BB'}
\label{W1}
\end{equation}
and
\begin{equation}
W^{B}=V_{(AA'),(BB')}- V_{AA'}\otimes I_{BB'}
\label{W2}
\end{equation}
with $V_{XY}$ is a ``swap'' or ``flip''
operator between space ${\cal H}_X$ and ${\cal H}_Y$
(see \cite{Werner0}. Evidently meanvalue
$\langle \langle W^{X} \rangle \rangle_{\varrho}
\equiv \langle W^{X} \langle_{\varrho \otimes \varrho}$
is positive iff the inequalities (\ref{q}), (\ref{alfa})
are satisfied for $X=A,B$ and $q=2$.
Thus, because $\langle \langle W^{X} \rangle \rangle_{\varrho}$
is (i) positive for all separable states $\varrho$
(ii) negative for some entangled states $\varrho$
(those that violate (\ref{q}), (\ref{alfa})) we propose
to call them {\it multicopy entanglement witnesses}.
In general it is likely that {\it multicopy observables}
like the above entanglement witnesses can be useful not only
in quantum information theory but quantum domain in general.

{\it Simple detection of entanglement by multicopy entanglement
witnesses.-}

The above scheme can be immediately generalised
to $n$-copies with one exeption - the multicopy generalisations
of (\ref{W1}), (\ref{W2}) are not hermitian, so to estimate
their ``mean values'' we need separate measurements of
hermitian and antihermitian parts.
The corresponding (nonhermitian) multiparticle operators
composed of bot parts could be  called
multicopy entanglement quasi-witnesses
as they detect entanglement but they are not hermitian.

\section{Remarks on state spectrum estimation}
It is remarkable that if we want to determine spectrum of
the state $\varrho$ defined on $m$ dimensional Hilbert space
than  {\it all we need} is the m-1 values $Tr(\varrho^{2})$,
$Tr(\varrho^{3})$, ..., $Tr(\varrho^{m})$.
This can be seen by realising that
$Tr(\varrho^{k})=(p_{1})^{k} + (p_2)^{k} + ...+ (p_m)^{k}$
where $p_{i}$ stand for spectrum of $\varrho$.
The efectiveness is determined by unique solution
because finite discrete random variable
is  determined by its first $m$ moments.

Thus we have the following

{\it Conclusion .- In order to determine the spectrum of
(completely unknown) state $\varrho$ it is enough to
determine the mean values of
$2m - 3$ observables: $V^{(2)}; V_{h}^{(3)}, V_{a}^{(3)};
V_{h}^{(4)}, V_{a}^{(4)}; ...;
V_{h}^{(m)}, V_{a}^{(m)}$
where $V_{a}^{(k)}, V_{h}^{(k)}$ are hermitian and aetyhermitian
parts of ``shift'' (or generalised swap) operation $V^{(k)}$. }

{\it Remark .- } This result is
complementary to what is known in literature
so far we have had either (i) to perform full tomography:
estimation of mean values of $m^2$ obseravbles, each
{\it via} single copy statistics
or (ii) asymptotic estimation of
Yang frames requiring estimation of sequence of
n-copy observables with $n$ approaching infinity.
The method (ii) was shown to be optimal (see\cite{Spectrum}).
It involves effectively
only $m$ observables
(because it requires estimation  of
probabilisties of results of $m$ output
observables) but requires possibility of collective
measurements on arbitrary number of copies
to get arbitrary good accuracy.
The former (i) requires only single copy per measurement
but $m^2$ observable are needed.
In present approach we see that only $2m-3$ observables we need if
collective measurements on
{\it finite} numbers of  $2$, $3$, ..., $m$ copies is possible.
In that sense it represents some kind of compromise between
to previous two methods (i) and (ii).

Finally note that the Renyi or Tsallis inequalities discussed in previous section
involve only spectra of system and subsystem
density matrices. Using the above spectrum estimation
for system and subsystem state one can check
the inequalities just puting the estimated spectra in them.

\section{Conclusions}
We have considered the possibility of transformation of getting
$n$-th power of the state $\varrho$ given $n$ copies of it.
We have shown that it is impossible.
We analysed possibility of structural
physical approximations (SPA)
of unphysical maps ones under the condition
of complete symmetry of approximation error $\Delta$.
We have pointed out that it is possible to approximate any linear
hermitian map in such a way. We have optimised
SPA of any nontrivial hermitian map
under the assumption of continuity
of shrinking factor $\gamma$.

On the other hand we have shown
that the natural tracepreserving approximation of (unphysical)
operation of the power of the state in the above
sense is impossible for power $2$ in the sense of SPA.
It was conjectured that this ``no go ''  property
it is true in general i. e. that no SPA of the transformation
$\varrho^{\otimes n} \rightarrow \varrho^{n}$ exists.

It has been pointed out that, however,
given two copies of $\varrho$ the nonlinear
function of the state defined by  $Tr(\varrho^{2})$
can be easily measurable. This leads to the physical
``twocopy observable'' measuring the Tsallis entropy.
This approach allows to consider notion of
multicopy observables and, in particular,
to leads to the definition of multicopy entanglement witnesses.
The exmples of the latter has been provided,
measuring degree of violation of separability
criterion based on entropic inequalities.
The method with $q=2$ can be partially
generalised to higher order Tsallis entropies which
can be estimated by means of only {\it two}
multicopy observables. Thus the corresponding separability
conditions can be checked by means of {\it pairs}
of hermitian observables.
Finally the existence of simple method of spectrum
estimation for unknown state has been pointed out
which requires collective measurements on small
number of copies. The number of needed estimated
parameters is $2 dim {\cal H}-3$ which is
less than  $(dim{ \cal H})^{2}-1$ required
in tomography. One can hope that the multicopy observables
idea together with physical interpretation of Tsallis
entropy in context of multicopy observables can
be useful not only for quantum entanglement theory but
also for quantum physics in general.

The author thanks Vladimir Buzek and Marek Czachor for
interesting discussions on nonlinearity in context of quantum mechanics.
He is grateful to Artur Ekert for pointing out inconsistency in
erlier version of this work and for helpful discussions.
Remarks of Joachim Domsta on classical probability
theory are also acknowledged.
The work is supported by Polish Committee for Scientific Research,
contract No. 2 P03B 103 16, and by the European
Union, project EQUIP, contract No. IST-1999-11053.

\section{Appendix}
Here according to well-known quantum mechanical procedure
known form Kraus (see \cite{Alicki})
we shall recall how any completely positive map $\Lambda$ satisfying
\begin{equation}
Tr[\Lambda(\varrho)] \leq 1
\label{trace}
\end{equation} can be
probabilistically implemented in the laboratory.
We know that
$\Lambda(\cdot)=\sum_{i=1}^{k}V_{i}(\cdot)V_{i}^{\dagger}$.
From (\ref{trace}) remembering that
$Tr[\sum_{i=1}^{k}V_{i}\varrho V_{i}^{\dagger}]=
Tr[\sum_{i=1}^{k}V_{i}^{\dagger}V_{i}\varrho]$.
we get that the positive operator
$A_{0}=\sum_{i=1}^{k}V_{i}^{\dagger}V_{i}$ has the spectrum
form the interval $[0,1]$. Thus we can define
$V_{0}=\sqrt{I-A_{0}}$ an then the extended completely positive map
$\Lambda'(\cdot)=\Lambda(\cdot)+ V_{0} (\cdot)V^{\dagger}=
\sum_{i=0}^{k}V_{i}(\cdot)V_{i}^{\dagger}$ is tracepreserving because
$Tr[\Lambda(\varrho)]=Tr[\sum_{i=0}^{k}V_{i}\varrho V_{i}^{\dagger}]=
Tr[\sum_{i=0}^{k}V_{i}^{\dagger}V_{i}\varrho]=
Tr[(A_{0}+I-A_{0}\varrho]=Tr(I\varrho)=Tr(\varrho)=1$.
But any tracepresering map can be implemented in lab
by interaction with some additional quantum system
(ancilla) and some von Neumann measurement on that system
with outputs $i=0, 1, ..., k$ (for description see
Ref. \cite{Alicki}). In that case the $i-th$
``event'' corresponds to the single map $V_{i} \cdot V_{i}^{\dagger}$.
It can be interpreted as ``producing''  unonrmalised state
$V_{i} \varrho V_{i}^{\dagger}$. Indeed though
its action results in normalised state
$\varrho_{i}=V_{i} \varrho V_{i}^{\dagger}/p_{i}$
this occurs only with probability $p_{i}=Tr(V_{i}\varrho V_{i}^{\dagger})$.
In this sense the original map $\Lambda$ can be implemented:
we apply some special von Neumann measurement on
the ancilla and keep the system if only the i-th event  with
$i \neq 0$ occurs. If the singled out event corresponding
to $i=0$ occurs we ``discard'' our system.
This gives the new state $\varrho'=\Lambda(\varrho)/p$ with
probability $p=Tr[\Lambda(\varrho)]$.

\end{document}